\documentclass[preprint,3p,10pt]{elsarticle}

\usepackage{graphicx}
\usepackage{amssymb}
\usepackage{lineno}

\usepackage{tikz}
\usetikzlibrary{shapes,arrows}

\usepackage{hyperref}
\hypersetup{
  colorlinks=true,
}

\journal{arxiv}

\begin{document}
\hypersetup{
  linkcolor=red,
  urlcolor=blue,
  citecolor=red
}

\begin{frontmatter}


\title{An Empirical Analysis of the Python Package Index (PyPI)}



\author{Ethan Bommarito}
\ead{bommare@umich.edu}
\author{Michael J Bommarito II}
\address{University of Michigan, Ann Arbor}
\address{Bommarito Consulting, LLC}
\ead{michael@bommaritollc.com}

\begin{abstract}
In this research, we provide a comprehensive empirical summary of the Python Package Repository, PyPI, including both package metadata and source code covering 178,592 packages, 1,745,744 releases, 76,997 contributors, and 156,816,750 import statements.  We provide counts and trends for packages, releases, dependencies, category classifications, licenses, and package imports, as well as authors, maintainers, and organizations.  As one of the largest and oldest software repositories as of publication, PyPI provides insight not just into the Python ecosystem today, but also trends in software development and licensing more broadly over time.  Within PyPI, we find that the growth of the repository has been robust under all measures, with a compound annual growth rate of 47\% for active packages, 39\% for new authors, and 61\% for new import statements over the last 15 years.  As with many similar social systems, we find a number of highly right-skewed distributions, including the distribution of releases per package, packages and releases per author, imports per package, and size per package and release.  However, we also find that most packages are contributed by single individuals, not multiple individuals or organizations.  The data, methods, and calculations herein provide an anchor for public discourse on PyPI and serve as a foundation for future research on the Python software ecosystem.
\end{abstract}

\begin{keyword}
software \sep software development \sep python \sep licensing \sep open source \sep dependency \sep complex system
\end{keyword}

\end{frontmatter}

% \linenumbers

\section{Introduction}
\label{S:introduction}

Since its first release in 1991, Python has established itself as a popular and widely-adopted general purpose programming language (\cite{bissyande2013}); in fact, the language continues to gain popularity at a high rate, having received the largest percentage increase for the 2018 annual TIOBE index (\cite{tiobe}).  One common explanation for this success is Python's rich ecosystem of libraries and applications.  Inspired by the TeX community's Comprehensive TeX Archive Network (CTAN)  and the Perl community's Comprehensive Perl Archive Network (CPAN), the Python community first laid the groundwork for its packaging ecosystem, PyPI, in 2001 and 2002 through PEP 241 and 301 (\cite{PEP241}, \cite{PEP301}).  While PyPI first began operating in 2003, its currently-available record begins in 2005 . During its nearly 15 years of operation, PyPI has grown to host over 175,000 packages, 1.7 million releases, and 2.3TB of package releases.

As such a large and longitudinal record, PyPI offers an empirical source of information about trends both specific to Python and general across software development.  There has been some recent research on such package repositories generally and limited research on Python and PyPI specifically (\cite{zheng2008analyzing}, \cite{samoladas2010survival}, \cite{orru2015curated}), \cite{decan2016topology}, \cite{decan2017empirical}, \cite{kikas2017structure}, \cite{malloy2017quantifying}, \cite{decan2019empirical}).  Many of these studies rely on samples that are hand-curated, very small relative to the overall package ecosystem, or that rely on incomplete or inaccurate metadata.  Most notably complete among extant literature is the work of Decan et al., which takes a comparative networks approach to nearly all major open source package repositories.  Their research has generally focused on dependency networks, however, glossing over other elements of the repositories.  With respect to Python, the best source of information is \cite{decan2016topology}, but this data source suffers from data coverage and quality issues, as it reflects only 20,522 packages and uses only requirements metadata that is often over- or under-inclusive relative to actual import statements in code.  As of 2019, the most comprehensive source of information on packages readily available is through Tidelift's libraries.io project \cite{tidelift2019}; however, libraries.io similarly relies exclusively on package metadata for license and dependency tracking, and, in some cases, has material errors in complex cases, such as that for psycopg2.

Motivated in part by \cite{decan2016topology} and experiences managing software development compliance, we set out to provide an broader characterization of PyPI, including both package metadata and package source and covering not just software dependencies at a point in time, but also authors, licenses, and other summary information over time.  This research is intended to provide a convenient reference for empirical claims regarding the Python ecosystem, and to provide anchor for a larger body of future research.

\section{Data and Methods}
\label{S:data}
PyPI, like many projects that span multiple decades, has seen numerous and substantial software and infrastructure changes since its first release and deployment in 2003.  The record presented and analyzed in this research is based on a sample retrieved from \small{\url{https://pypi.org}} in May of 2019, using the recently-released Warehouse API endpoints at \small{\url{https://pypi.org/simple/}} and \small{\url{https://pypi.org/pypi/p/json/}}.  The retrieval procedure is described below in \ref{S:data_retrieval_procedure}.  This sample spans nearly 15 years from March 2005 through May 2019.

\subsection{Data Retrieval Procedure}
\label{S:data_retrieval_procedure}
\begin{enumerate}
    \item Retrieve the list of all packages from \small{\url{https://pypi.org/simple/}}
    \item For each package $P$,
    \begin{enumerate}
        \item Retrieve package metadata from \small{\url{https://pypi.org/pypi/P/json/}}
        \item Parse and store package metadata
        \item For each release $R$ in package $P$,
        \begin{enumerate}
            \item Parse and store release metadata from package JSON
            \item For each source package $S$ in release $R$ (package type$=$``sdist''),
            \begin{enumerate}
                \item Retrieve the source package archive from the primary \small{\url{pypi.org}} mirror
                \item Parse the sdist package to identify and store Python dependencies (see \ref{S:dependency_identification_procedure} below)
            \end{enumerate}
        \end{enumerate}
    \end{enumerate}
\end{enumerate}

\subsection{Dependency Identification Procedure}
\label{S:dependency_identification_procedure}

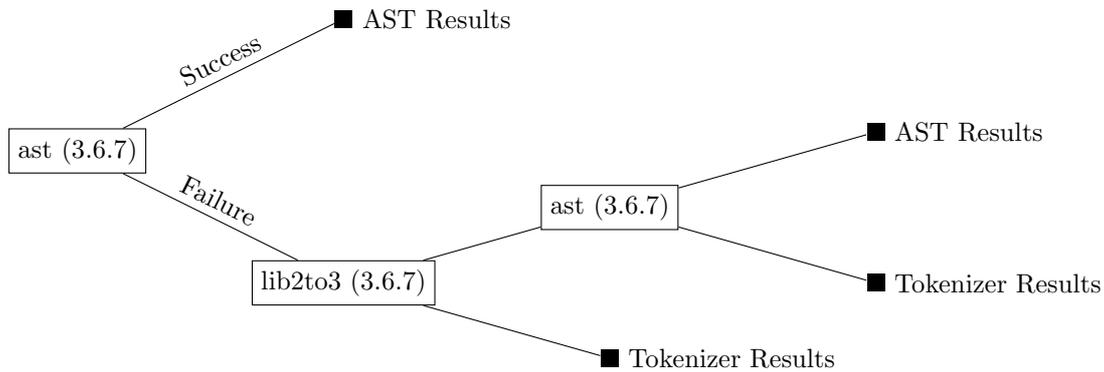
\begin{figure}[ht!]
   \centering
    
    % Set the overall layout of the tree
    \tikzstyle{level 1}=[level distance=3.5cm, sibling distance=3.5cm]
    \tikzstyle{level 2}=[level distance=3.5cm, sibling distance=2cm]
    
    % Define styles for bags and leafs
    \tikzstyle{bag} = [draw]
    \tikzstyle{end} = [rectangle,fill]

    \begin{tikzpicture}[grow=right,sloped]
    \node[bag] {ast (3.6.7)}
		child {
			node[bag] {lib2to3 (3.6.7)}
			child {
				node[end,label=right:Tokenizer Results] {}
				edge from parent
			}
			child {
				node[bag] {ast (3.6.7)}
				child {
					node[end,label=right:Tokenizer Results] {}
					edge from parent
				}
				child {
					node[end,label=right:AST Results] {}
					edge from parent
				}
				edge from parent
			}
			edge from parent
			node[above] {Failure}
		}
		child {
            node[end,label=right:AST Results] {}
            edge from parent
            node[above] {Success}
        }
    ;
    \end{tikzpicture}
    \caption{Dependency Identification Flowchart}
    \label{fig:dependency_identification_flowchart}
\end{figure}

Once all package metadata and release source packages are retrieved, we next identify package dependencies.  Unlike most extant research, we analyze actual source code files instead of relying solely on package metadata or requirements file.  As many software developers are aware, package repository metadata often omits dependencies required for some or all package functionality.  This may be accidental, as often occurs in release management of complex or immature software projects, or may be purposeful, such as for licensing or package size considerations.  A detailed description of our dependency identification procedure is provided below, including a flowchart in figure \ref{fig:dependency_identification_flowchart}.

All source packages in all releases are attempted so long as they are archives with one of the following extensions: .zip, .egg, .tar, .tar.gz, .tgz, .tar.bz2, .tbz.  For each source package $S$, each object $O$ in the archive with extension .py is parsed using a fallback approach.  First, within a Python 3.6.7 interpreter on Ubuntu 18.04, we attempt to parse the source object into an AST tree using the \textsc{ast.parse method} (\cite{python_ast}).  If this ast method succeeds, then we walk the tree and extract information from all ast.Import and ast.ImportFrom nodes.  If this ast method fails, then we next assume that the source object is from a backwards-incompatible release of the Python interpreter.  We attempt to correct this using a lib2to3.refactor.RefactoringTool class instantiated with the standard lib2to3.fixes fix set (\cite{python_lib2to3}). If RefactoringTool succeeds, we attempt again to parse the refactored source using ast.parse and return ast.Import and ast.ImportFrom nodes from the tree.  If either RefactoringTool fails or ast.parse fails on the refactored source, then we fall back on the final token parser, which searches all lines for \textsc{import X} or \textsc{from X import Y} statements.

\subsection{License Identification and Normalization Procedure}
\label{S:license_identification_procedure}
Package licensing is not a simple topic; packages may split or combine, change ownership, change license, offer multiple licenses for the entire package, license subsets of a package separately, or vendor other packages with other licenses.  From a legal perspective, the proper unit of analysis may be context-dependent and difficult to identify generally; is it a parse tree or subset thereof, line of code, ``file'', package, or release?  Is it a package at a point in time or over its entire history? In the context of an interpreted language like Python, it becomes even more complicated, as the relationship between source code and executed instructions is far more complex than in most ``fully-compiled'' languages.

In this research, we attempt to determine package licensing by assessing three simply observable facts from the most recent package info and release source code.  First, we look at the standard license metadata field provided to setuptools and parsed by PyPI.  We normalize this raw license string, which may be a license name, a URL or path, or complete license text, to map to a license name, license version, and license family.  For example, values such as \textit{New BSD}, \textit{BSD 3}, and \textit{BSD 3 Clause License} are all mapped to a standard record: (name=BSD, version=3-Clause, family=BSD).  This normalization is based on a mapping of over 500 license names, canonical license URLs (e.g., opensource.org or gnu.org), and common license phrases across over 5,000 variations of this field.  

Second, we review package metadata to see if the contributor references a GitHub repository as its home page, as over 116,000 package in PyPI do so.  For such packages, we then check to see whether the contributor specified a path in the license metadata, and retrieves this path from GitHub if present; if no license metadata field was set, then we search for default file names on GitHub such as LICENSE, LICENSE.txt, LICENSE.md, or LICENSE.rst.  If located, we search these license files for key phrases as when full license source is provided in the license metadata field.  Again, the end result is a mapping to license name, license version, and license family

Finally, we look for one or more license classifications assigned by the package author on PyPI.  While these classifications cover approximately 102,000 packages, the license taxonomy and validation on PyPI result in data quality issues.  For example, PyPI does not distinguish between any variations of the BSD License, and also makes available three different options for the GPL License, Version 3.  As an example, the alvi package has assigned 9 License classifications, including DFSG Approved, Free for Home Use, Freeware, and MIT License, whereas over 70,000 packages list no License classifications.

\begin{figure}[ht!]
    \centering
    
    % Set the overall layout of the tree
    \tikzstyle{level 1}=[level distance=4cm, sibling distance=4cm]
    \tikzstyle{level 2}=[level distance=3cm, sibling distance=3cm]
    
    % Define styles for bags and leafs
    \tikzstyle{bag} = [draw]
    \tikzstyle{end} = [rectangle,fill]

    \begin{tikzpicture}[grow=right,sloped]
    \node[bag] {setuptools:license}
		child {
			node[bag] {LICENSE file}
			child {
				node[bag] {setuptools:classifiers}
				child {
					node[end,label=right:Unknown] {}
					edge from parent
				}
				child {
					node[end,label=right:Assigned] {}
					edge from parent
				}
				edge from parent
			}
			child {
				node[end,label=right:Assigned] {}
				edge from parent
			}
			edge from parent
			node[above] {Failure}
		}
		child {
            node[end,label=right:Assigned] {}
            edge from parent
            node[above] {Success}
        }
    ;
    \end{tikzpicture}
    \caption{License Identification Flowchart}
    \label{fig:license_identification_flowchart}
\end{figure}
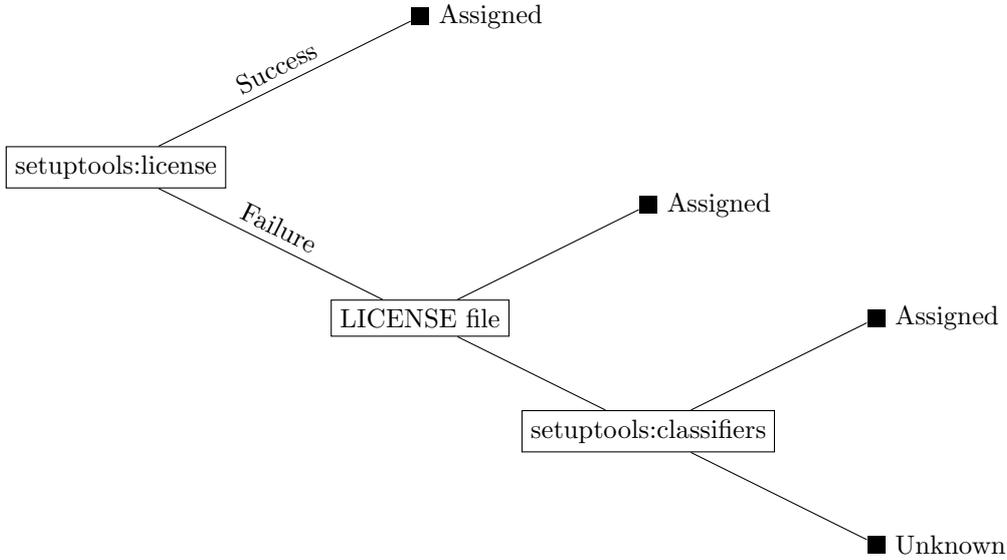

Based on these three sources of license information, we implement a fallback strategy similar to that used above for dependency parsing.  This strategy is visualized in the flowchart in Figure \ref{fig:license_identification_flowchart}.  Our approach, like all extant sources we have identified, does not address multiple licensing or sublicensing, and fails to identify a license when non-standard language or disclosure locations are used.  In future work, we intend to detail a more complete approach to license identification that captures the true legal complexity of software licensing.

These methods are all implemented in Python itself, and the results are stored in a Postgres SQL database for ease of automation, replication, and querying.  The authors have open-sourced the code and data required to replicate these results at \url{https://github.com/bommarito-consulting/pypi-research-data}, and if enough interest is generated, may periodically update this repository to reflect future results.

\section{Results}
\label{S:results}

This paper is intended to provide a high-level empirical overview of the PyPI ecosystem as of May 2019; as such, our results largely consist of raw counts and proportions by categorical dimension or over time.  While a limitless number of more detailed causal or normative questions could be asked and answered, we limit the scope of this paper to provide simple, direct statistics.  Please see Section \ref{S:conclusion} below for discussion of related in-progress and future works.

Table \ref{tab:summary_statistics} demonstrates the size of the PyPI ecosystem in aggregate across its most fundamental measures.  As the fourth largest software development repository based on \cite{debill2019}, PyPI contains over 178,000 packages and over 1.7 million individual package releases released by more than 76,000 unique authors.  From a metadata and code perspective, there are nearly 1 million assigned package classifications and over 156 million imports evidenced within the 2.4TB of release packages.

\begin{table}[ht]
    \centering
    \begin{tabular}{r|r}
    \hline
    \textbf{Statistic} & \textbf{Value}\\
    \hline
    Number of packages & 178,952\\
    Number of releases & 1,745,744\\
    Number of package classifications & 947,896\\
    Number of authors & 76,997\\
    Number of maintainers & 3,047\\
    Number of licenses (raw) & 4,610\\
    Number of imports & 156,816,750\\
    \hline
    \end{tabular}
    \caption{Summary of PyPI ecosystem statistics (May 2019)}
    \label{tab:summary_statistics}
\end{table}

While PyPI is clearly ``large'' by most measures, there are a number of related questions.  For example, is PyPI getting larger at an increasing rate of packages, releases, or authors - i.e., is the ecosystem accelerating or decelerating?  Is the ecosystem growing through one-off packages that are quickly abandoned, or is the number of actively maintained packages growing too? Table \ref{tab:stats_by_year_package_release_author} provides insight to these questions by documenting the number of new packages, active packages, new releases, and new authors by year.  New packages are calculated from year of first release upload time, active packages are calculated by unique packages with at least one release uploaded per year, new releases are calculated from year of upload time, and new authors are calculated by year of first uploaded release across all authored packages.

Table \ref{tab:stats_by_year_package_release_author} clearly demonstrates that the PyPI ecosystem is accelerating across all measures; Table \ref{tab:stats_pypi_cagr} calculates the actual compound annual growth rate for full years 2006 through 2018 for each measure above.  All measures show sustained double-digit growth rates over the 13 year period, with a headline rate of 51\% additional releases per year and 39\% new authors per year.

\begin{table}[ht!]
    \centering
    \begin{tabular}{c|r|r|r|r}
    \hline
    \textbf{Year} & \textbf{New Packages} & \textbf{Active Packages} & \textbf{New Releases} & \textbf{Authors} \\
    \hline
    2005 &                  96 & 96 & 389 &                  68 \\
    2006 &                 367 & 420 & 2,324 &                 216 \\
    2007 &                 876 & 1,047 & 5,301 &                 341 \\
    2008 &                1,702 & 2,223 & 10,923 &                 637 \\
    2009 &                2,559 & 3,727 & 17,426 &                 974 \\
    2010 &                3,522 & 5,454 & 23,474 &                1,455 \\
    2011 &                4,840 & 7,479 & 31,107 &                1,996 \\
    2012 &                7,235 & 11,047 & 47,377 &                3,062 \\
    2013 &               10,438 & 16,267 & 74,003 &                4,633 \\
    2014 &               13,352 & 21,555 & 112,994 &                6,016 \\
    2015 &               17,355 & 28,498 & 173,437 &                7,742 \\
    2016 &               21,849 & 36,557 & 253,262 &               10,293 \\
    2017 &               29,905 & 48,223 & 372,034 &               12,374 \\
    2018 &               39,351 & 64,628 & 502,029 &               16,064 \\
    2019 &               16,873 & 18,919 & 199,664 &                6,938 \\
    \hline
    \end{tabular}
    \caption{Number of new packages, active packages, new releases, and new authors on PyPI by year; 2019 is a partial year through May 15}
    \label{tab:stats_by_year_package_release_author}
\end{table}

\begin{table}[ht!]
    \centering
    \begin{tabular}{r|r}
        \hline
        \textbf{Measure} & \textbf{CAGR (\%)} \\
        \hline
        New Packages    &  43.28 \\
        Active Packages &  47.31 \\
        New Releases    &  51.21 \\
        New Authors     &  39.30 \\
        \hline
    \end{tabular}
    \caption{Compound Annual Growth Rate (CAGR) for new packages, active packages, new releases, and new authors on PyPI}
    \label{tab:stats_pypi_cagr}
\end{table}

Next, we examine how the ecosystem's size in releases is distributed across packages and how these packages behave from a release timing perspective.  Table \ref{tab:stats_releases_by_package} documents the mean, standard deviation, and common percentiles for number of releases by package and the distribution of days between releases, i.e., the inter-release timing distribution.  

These statistics paint a commonly observed pattern of right-skewed counts; a small number of packages accounts for a large percentage of releases.  For example, the 100 top packages by releases account for 3.79\% of all releases, despite representing only 0.05\% of packages. To quantify the degree of skew, we calculate the Gini coefficient for the number of releases per package, finding a value of 0.6274; the Gini coefficient is a simple and well-known measure of ``equality'' of such distributions (\cite{relative1979}).  For comparison, the distribution of global wealth is estimated by the UN to be between 0.61 and 0.68 (\cite{wealth2010}).  The difference between these releases rates can be explained not just by package activity or duration, but also by varying release ``philosphies'' and the presence of continuous integration (CI) processes.  For example, many of the largest packages by count ``release'' to PyPI every night, or in the extreme case, on every git commit or merge to a specific branch.  Such packages will clearly outpace those with ``traditional'' approaches like monthly, quarterly, or ``major'' releases. 

\begin{table}[ht!]
    \centering
    \begin{tabular}{r|r|r}
    \hline
    \textbf{Statistic} & \textbf{Releases per Package} & \textbf{Inter-Release Timing (days)}\\
    \hline
    Mean  &       6.93 & 65.38\\
    Standard Deviation   &      19.46 & 144.89\\
    Minimum   &       1 & 0 \\
    25th Percentile  &       1 & 1.18\\
    50th Percentile   &       3 & 16.37\\
    75th Percentile   &       7 & 68.12 \\
    Max   &    4,333 & 4,369\\
    \hline
    \end{tabular}
    \caption{Descriptive statistics for distribution of number of release per package}
    \label{tab:stats_releases_by_package}
\end{table}

We continue this line of analysis by next reviewing the distribution of packages and releases by author.  Table \ref{tab:stats_by_author} documents the mean, standard deviation, and common percentiles for this distribution, demonstrating a similarly right-skewed property.  We also calculate the Gini coefficients, finding a value of 0.4595 for the number of packages per author and a value of 0.7331 for number of releases per author.  In general, we find that most authors contribute a single package with a small number of releases, and as the Active Package counts in Table \ref{tab:stats_by_year_package_release_author} demonstrate, nearly two in every three packages went without a release during 2018.  Some such packages may be mature and ``complete,'' requiring no further releases by the author; others, however, may reflect the difficulty of sustaining effort and attention once the ideation and initial development stages have passed.  Notably, many of the largest contributors fall into two categories - framework maintainers such as Zope or Odoo who use PyPI like an ``app store'' for their application ecosystems and automated systems such as the typosquatting security efforts of William Bengtson.

\begin{table}[ht!]
    \centering
    \begin{tabular}{r|r|r}
    \hline
    \textbf{Statistic} & \textbf{Packages per Author} & \textbf{Releases per Authors}\\
    \hline
    Mean  &       2.15 & 15.13 \\
    Standard Deviation   &      17.41 & 125.47 \\
    Minimum   &       1 & 1 \\
    25th Percentile  &       1 & 2 \\
    50th Percentile   &       1 & 4 \\
    75th Percentile   &       2 & 12 \\
    Max   &    4,483 & 28,647 \\
    \hline
    \end{tabular}
    \caption{Descriptive statistics for distribution of number of packages and releases per author}
    \label{tab:stats_by_author}
\end{table}

Digging further into authorship, we next examine the \textit{types} of authorship evident through metadata - namely, what proportion of packages have multiple listed authors and what proportion of packages are published by an organization like for-profit or not-for-profit corporations.  We determine whether a package has multiple authorship by checking for a comma, the substring `` and '' or the phrase ``et al.''  We determine whether a package has organization authorship by checking for 13 common entity abbreviations and types (e.g., Inc., Ltd. SRL, GmbH) and 9 common tokens (e.g., Foundation, Developers, Lab).  Table \ref{tab:stats_by_author_type} documents the results, indicating that approximately 7.77\% of packages are authored by organizations and 7.01\% of packages are authored by multiple listed authors.  In future work, we intend to examine the distribution of authors by commonly discussed demographics such as inferred language, gender, age, and geography.

\begin{table}[ht!]
    \centering
    \begin{tabular}{r|r|r}
    \hline
    \textbf{Statistic} & \textbf{Author String} & \textbf{Packages}\\
    \hline
    Organization & 0.054911 & 0.077691 \\
    Multiple Authors & 0.083016 & 0.070052 \\
    \hline
    \end{tabular}
    \caption{Descriptive statistics for proportion of author strings and packages that appear to be organizations or multiple authors}
    \label{tab:stats_by_author_type}
\end{table}

Next, we examine how much storage space packages and releases generally use on PyPI.  Table \ref{tab:size_stats_by_release} documents the distribution of size in kilobytes by packages and releases.  We find that most packages and releases are quite small, with a median release size of 22.6KiB and a median package size of 40.0KiB.  However, right-skew again appears, with the largest releases weighing in at nearly 600 megabytes and the largest packages using nearly 175 gigabytes.  Together, the four ``deep learning'' packages tf-nightly, mxnet-cu100mkl, mxnet-cu100, and tf-nightly-gpu use approximately 500 gigabytes of PyPI storage - nearly 25\% of all PyPI storage.  The Gini coefficient of the distribution of package sizes summarizes the extreme inequality with a value of 0.9902. While the convenience of pip and tight model and code integration have led to such large releases and packages, the community may in the future consider whether alternative methods of distribution or storage-based costing may be more appropriate for such packages.

\begin{table}[ht!]
    \centering
    \begin{tabular}{r|r|r}
    \hline
    \textbf{Statistic} & \textbf{Size by Release (KiB)} & \textbf{Size by Package (KiB)}\\
    \hline
    Mean &    1,606.55 & 16,466.79 \\
    Standard Deviation  &   14,971.50 & 809,136.51 \\
    Minimum  &       0.00 & 0.02 \\
    25th Percentile  &       7.50 & 10.04 \\
    50th Percentile  &      22.60 & 40.00 \\
    75th Percentile  &     128.16 & 216.28 \\
    Maximum  &  592,437.94 & 175,763,749.00 \\
    \hline
    \end{tabular}
    \caption{Descriptive statistics for distribution of size by release and package}
    \label{tab:size_stats_by_release}
\end{table}

Next, we shift our attention to the relative frequency of licenses assigned by authors, as inferred with the process describe in Section \ref{S:license_identification_procedure} above.  Table \ref{tab:stats_by_license_family} documents the distribution of packages by license family, including the proportion of each license family out of the total.  Firstly, over half of all packages on PyPI are licensed under permissive terms such as MIT, BSD, or Apache.  Secondly, only 16.4\% of PyPI packages are themselves explicitly licensed under a GPL family license, including LGPL, GPL, and AGPL variants.  Thirdly, over 25\% of packages are packaged and contributed to PyPI without clear license metadata.  The community may consider whether new contributions to PyPI should be required to provide more explicit license metadata in order to better protect the ecosystem from cascading licensing issues.

\begin{table}[ht!]
    \centering
    \begin{tabular}{r|r|r}
    \hline
    \textbf{License Family} & \textbf{Number of Packages} & \textbf{Proportion}\\
    \hline
    MIT           &  60,945 &  0.340566 \\
    Unknown       &  48,742 &  0.272375 \\
    GPL           &  29,403 &  0.164307 \\
    BSD           &  20,094 &  0.112287 \\
    Apache        &  15,004 &  0.083844 \\
    Public Domain &   1,194 &  0.006672 \\
    Zope          &   1,150 &  0.006426 \\
    ISC           &    719 &  0.004018 \\
    MPL           &    712 &  0.003979 \\
    PSFL          &    524 &  0.002928 \\
    Proprietary   &    190 &  0.001062 \\
    CC            &    178 &  0.000995 \\
    CeCILL        &     72 &  0.000402 \\
    zlib          &     25 &  0.000140 \\
    \hline
    \end{tabular}
    \caption{Number of packages and proportion of total by license}
    \label{tab:stats_by_license_family}
\end{table}

As its licenses and variants demonstrate the most range, we next examine the GPL family of licenses in particular.  Table \ref{tab:stats_by_license_gpl_family} documents the distribution of packages by license family, including the proportion of each license family out of the total.  Out of all GPL-family-licensed packages, we find that the LGPL is used nearly 12\% of the time, the AGPL is used approximately 21\% of the time, and the standard GPL is used approximately 68\% of the time.  Unfortunately, over one-third of these GPL license assignments do not specify a version (i.e., 2 or 3).

\begin{table}[ht!]
    \centering
    \begin{tabular}{r|r|r|r}
    \hline
    \textbf{License} & \textbf{Version} & \textbf{Count} & \textbf{Proportion}\\
    \hline
AGPL & 3 &   5,642 &  0.191885 \\
     & Unknown &    683 &  0.023229 \\
GPL & 2 &   2,428 &  0.082577 \\
     & 2.1 &     69 &  0.002347 \\
     & 3 &   84,37 &  0.286944 \\
     & Unknown &   8508 &  0.289358 \\
LGPL & 2 &    122 &  0.004149 \\
     & 2.1 &     90 &  0.003061 \\
     & 3 &   1,501 &  0.051049 \\
     & Unknown &   1,923 &  0.065401 \\
    \hline
    \end{tabular}
    \caption{Number of packages and proportion of total by version for GPL family licenses}
    \label{tab:stats_by_license_gpl_family}
\end{table}

There has been limited empirical work on the relative frequency of license adoption and violations until recently (\cite{license2015}, \cite{machine2017}, \cite{adopt2015}, \cite{open2019}, \cite{almeida2019}, \cite{analysis2017}), and almost none of the extant research has investigated the problem from a dependency graph perspective. While further treatment and discussion of licensing within PyPI, including identification of license transitions, unauthorized software re-use, and invalid chains of dependency licensing, are outside of the scope of this research, the authors are actively developing a system for this purpose and analysis, to be published in future work. 

We next examine the most common classifications assigned to packages by authors.  The PyPI classification taxonomy is designed for information retrieval purposes, and its primary use case is to assist those searching for software by topic, development status, intended audience, or operating system.  There are nearly 1 million assigned classifications on PyPI as of May 2019.

We first examine the Development Status classification category.  Table \ref{tab:stats_by_development_status} documents the distribution of assigned classifications by development status classification, including the proportion of each status out of the total.  Notably, barely 25\% of all packages classify themselves as Production/Stable or Mature; the modal classification is Beta, followed in frequency by Alpha.  While it is likely based on the ratio of active packages to cumulative new packages in Table \ref{tab:stats_by_year_package_release_author} and the distribution of maturity statuses in Table \ref{tab:stats_by_development_status} that tens of thousands of packages are inactive or abandoned, only 314 packages have assigned the Inactive label on PyPI to alert potential users.

\begin{table}[ht!]
    \centering
    \begin{tabular}{l|r|r}
    \hline
    \textbf{Development Status} & \textbf{Count} & \textbf{Proportion} \\
    \hline
    1 - Planning          &   2,619 &  0.031979 \\
    2 - Pre-Alpha         &   6,409 &  0.078256 \\
    3 - Alpha             &  22,686 &  0.277003 \\
    4 - Beta              &  28,960 &  0.353611 \\
    5 - Production/Stable &  20,425 &  0.249396 \\
    6 - Mature            &    485 &  0.005922 \\
    7 - Inactive          &    314 &  0.003834 \\
    \hline
    \end{tabular}
    \caption{Number of assigned classifications and proportion of total by development status}
    \label{tab:stats_by_development_status}
\end{table}

Next, we examine the Intended Audience classification.  This category is intended to help segment potential users of software, e.g., to separate ``finished'' software for end users from ``intermediate'' software intended to be used by developers as part of a larger ``finished'' application.  Table \ref{tab:stats_by_audience} documents the distribution of assigned classifications by audience classification, including the proportion of each status out of the total.  Unsurprisingly, the majority of PyPI packages with labels are intended to be used by Developers, with 66\% packages receiving this label.  Next most frequent is the Science/Research category; given the popularity of Python among academics and industrial researchers and the large ecosystems fostered by packages such as numpy, scipy, astropy, and Bioypthon, this is also unsurprising.  A small number of other audiences including System Administrators, End Users/Desktop, Information Technology, and Education have over 1\% of assigned labels, but all other audiences have \textit{de minimis} representation.  

\begin{table}[ht!]
    \centering
    \begin{tabular}{r|r|r}
    \hline
    \textbf{Audience} & \textbf{Count} & \textbf{Proportion} \\
    \hline
    Developers                       &  69,682 &  0.668188 \\
    Science/Research                 &  12,020 &  0.115261 \\
    System Administrators            &   7,328 &  0.070269 \\
    End Users/Desktop                &   4,557 &  0.043698 \\
    Information Technology           &   4,132 &  0.039622 \\
    Education                        &   2,759 &  0.026456 \\
    Financial and Insurance Industry &    872 &  0.008362 \\
    Other Audience                   &    761 &  0.007297 \\
    Telecommunications Industry      &    529 &  0.005073 \\
    Healthcare Industry              &    526 &  0.005044 \\
    Legal Industry                   &    426 &  0.004085 \\
    Manufacturing                    &    383 &  0.003673 \\
    Customer Service                 &    222 &  0.002129 \\
    Religion                         &     88 &  0.000844 \\
    \hline
    \end{tabular}
    \caption{Number of assigned classifications and proportion of total by intended audience}
    \label{tab:stats_by_audience}
\end{table}

Next, we examine how packages vary in their compatibility with various operating systems.  Table \ref{tab:stats_by_os} documents the distribution of assigned classifications for the Operating System classification, including the proportion of each status out of the total. The OS Independent section is the most common classification by far, making up over half of the pool by itself; this is not surprising given that one of Python's perceived strengths is its cross-platform interpreted model and high degree of platform abstraction in standard libraries.  Among non-OS-independent labels, both POSIX and Mac categories outpace the Microsoft category, which accounts for only approximately 7.9\% of all packages.  Given Microsoft's recent push in 2019 to make Python more accessible and better supported on Windows 10, these relative ranks may be set to change in the future.

\begin{table}[ht!]
    \centering
    \begin{tabular}{r|r|r}
    \hline
    \textbf{Operating System} & \textbf{Count} & \textbf{Proportion} \\
    \hline
    OS Independent    &  47,053 &  0.577090 \\
    POSIX             &  16,036 &  0.196676 \\
    MacOS             &   7,240 &  0.088796 \\
    Microsoft         &   6,410 &  0.078617 \\
    Unix              &   4,573 &  0.056086 \\
    Android           &     93 &  0.001141 \\
    Other OS          &     49 &  0.000601 \\
    iOS               &     44 &  0.000540 \\
    OS/2              &     17 &  0.000208 \\
    BeOS              &     11 &  0.000135 \\
    PDA Systems       &      5 &  0.000061 \\
    PalmOS            &      4 &  0.000049 \\
    \hline
    \end{tabular}
    \caption{Number of assigned classifications and proportion of total by operating system}
    \label{tab:stats_by_os}
\end{table}

One of the Python ecosystem's most notable perceived strengths has been the maturity and flexibility of its web application and content management system (CMS) development frameworks, such as Django or Plone, and PyPI facilitates the management of communities for these frameworks through its Framework classification.  Table \ref{tab:stats_by_framework} documents the distribution of assigned classifications for this Framework classification, including the proportion of each status out of the total, for the top 20 records by count. Django is by far the most common of these frameworks, totalling nearly half of all assigned classifications by itself. Plone, Odoo, and Zope together make up another third of the assigned classifications, with no other framework totalling more than 2\%.

\begin{table}[ht!]
    \centering
    \begin{tabular}{r|r|r}
    \hline
    \textbf{Framework} & \textbf{Count} & \textbf{Proportion} \\
    \hline
    Django            &  17,337 &  0.472205 \\
    Plone             &   5,983 &  0.162958 \\
    Odoo              &   5,625 &  0.153207 \\
    Zope3             &    942 &  0.025657 \\
    Zope2             &    931 &  0.025357 \\
    Flask             &    662 &  0.018031 \\
    Tryton            &    556 &  0.015144 \\
    Pyramid           &    527 &  0.014354 \\
    Buildout          &    502 &  0.013673 \\
    Pytest            &    368 &  0.010023 \\
    IPython           &    330 &  0.008988 \\
    Twisted           &    324 &  0.008825 \\
    AsyncIO           &    321 &  0.008743 \\
    Sphinx            &    312 &  0.008498 \\
    Pylons            &    267 &  0.007272 \\
    TurboGears        &    187 &  0.005093 \\
    Jupyter           &    168 &  0.004576 \\
    Paste             &    151 &  0.004113 \\
    Bob               &    139 &  0.003786 \\
    Trac              &    107 &  0.002914 \\
    \hline
    \end{tabular}
    \caption{Number of assigned classifications and proportion of total by framework}
    \label{tab:stats_by_framework}
\end{table}

While most of the classifications reviewed thus far address technical aspects of a package, the Topic classification addresses the functional purpose or nature of the package; as such, it is one of the most commonly-used classifications on PyPI.  Table \ref{tab:stats_by_topic} documents the distribution of assigned classifications by Topic classification, including the proportion of each status out of the total. At just under fifty thousand instances or 37.8\% of assigned labels, Software Development is the most common of these topics, mirroring the Intended Audience classification.  Software Development is over twice as common as the next two most common topics, Internet and Scientific/Engineering, which together make up another 28\% of assigned labels. The remaining third of Topic labels have single-digit or smaller penetration, such as Printing, Adaptive Technologies, Sociology and Religion.

\begin{table}[ht!]
    \centering
    \begin{tabular}{r|r|r}
    \hline
    \textbf{Topic} & \textbf{Count} & \textbf{Proportion} \\
    \hline
    Software Development   &  48,887 &  0.377710 \\
    Internet               &  19,883 &  0.153620 \\
    Scientific/Engineering &  16,881 &  0.130426 \\
    Utilities              &  12,106 &  0.093533 \\
    System                 &   9,319 &  0.072000 \\
    Text Processing        &   4,445 &  0.034343 \\
    Multimedia             &   3,422 &  0.026439 \\
    Database               &   2,615 &  0.020204 \\
    Communications         &   2,518 &  0.019455 \\
    Office/Business        &   2,110 &  0.016302 \\
    Security               &   1,941 &  0.014997 \\
    Documentation          &   1,062 &  0.008205 \\
    Games/Entertainment    &    951 &  0.007348 \\
    Education              &    932 &  0.007201 \\
    Terminals              &    583 &  0.004504 \\
    Home Automation        &    421 &  0.003253 \\
    Desktop Environment    &    345 &  0.002666 \\
    Text Editors           &    290 &  0.002241 \\
    Other/Nonlisted Topic  &    227 &  0.001754 \\
    Artistic Software      &    204 &  0.001576 \\
    Printing               &    111 &  0.000858 \\
    Adaptive Technologies  &     87 &  0.000672 \\
    Sociology              &     53 &  0.000409 \\
    Religion               &     37 &  0.000286 \\
    \hline
    \end{tabular}
    \caption{Number of assigned classifications and proportion of total by topic}
    \label{tab:stats_by_topic}
\end{table}

PyPI's classification taxonomy supports hierarchical labels through the use of ``::'' delimiters, and the Topic schema features many additional subtopics.  Table \ref{tab:stats_by_topic_software} documents the distribution of assigned classifications by the Software Development subtopic of the Topic classification, i.e., labels matching ``Software Development :: *'', including the proportion of each status out of the total. A number of these labels, especially the most common Libraries label accounting for nearly two-thirds of assigned labels, appear relatively redundant or vague from an information retrieval perspective.  However, some labels, such as Build Tools, Testing, or Documentation, provide more specific and valuable information.  Given that PyPI's users are typically software developers, the community might consider whether this taxonomy could be improved or updated to facilitate better package discovery.

\begin{table}[ht!]
    \centering
    \begin{tabular}{r|r|r}
    \hline
    \textbf{Topic} & \textbf{Count} & \textbf{Proportion} \\
    \hline
    Libraries :: Python Modules         &  22,700 &  0.524734 \\
    Libraries                           &   6,781 &  0.156750 \\
    Build Tools                         &   4,056 &  0.093759 \\
    Testing                             &   2,489 &  0.057536 \\
    Libraries :: Application Frameworks &   1,792 &  0.041424 \\
    Quality Assurance                   &    969 &  0.022399 \\
    User Interfaces                     &    681 &  0.015742 \\
    Documentation                       &    543 &  0.012552 \\
    Code Generators                     &    539 &  0.012460 \\
    Version Control                     &    434 &  0.010032 \\
    Embedded Systems                    &    264 &  0.006103 \\
    Debuggers                           &    248 &  0.005733 \\
    Widget Sets                         &    220 &  0.005086 \\
    Compilers                           &    215 &  0.004970 \\
    Interpreters                        &    170 &  0.003930 \\
    Bug Tracking                        &    161 &  0.003722 \\
    Internationalization                &    142 &  0.003282 \\
    Object Brokering                    &    135 &  0.003121 \\
    Localization                        &    127 &  0.002936 \\
    Version Control :: Git              &    115 &  0.002658 \\
    \hline
    \end{tabular}
    \caption{Number of assigned classifications and proportion of total by Software Development subtopic}
    \label{tab:stats_by_topic_software}
\end{table}

Similarly, researchers in academic and industry contexts frequently apply subtopic labels under the Scientific/Engineering topic.  Table \ref{tab:stats_by_topic_scientific} documents the distribution of assigned classifications by the Scientific/Engineering subtopic of the Topic classification, including the proportion of each status out of the total. Bio-Informatics is the most commonly assigned topic at 17\%, but is closely followed by the three subtopics Artificial Intelligence, Mathematics, and Information Analysis, which are all around 13\%.  Notably, these four subtopics make up over half of the entire pool, and most likely correspond to the large communities like numpy and biopython discussed above.

\begin{table}[ht!]
    \centering
    \begin{tabular}{r|r|r}
    \hline
    \textbf{Topic} & \textbf{Count} & \textbf{Proportion} \\
    \hline
    Bio-Informatics                      &   2,050 &  0.170068 \\
    Artificial Intelligence              &   1,614 &  0.133897 \\
    Mathematics                          &   1,582 &  0.131243 \\
    Information Analysis                 &   1,574 &  0.130579 \\
    Physics                              &   1,253 &  0.103949 \\
    Visualization                        &    979 &  0.081218 \\
    Astronomy                            &    624 &  0.051767 \\
    GIS                                  &    578 &  0.047951 \\
    Chemistry                            &    444 &  0.036834 \\
    Medical Science Apps.                &    415 &  0.034428 \\
    Image Recognition                    &    239 &  0.019827 \\
    Atmospheric Science                  &    236 &  0.019579 \\
    Human Machine Interfaces             &    187 &  0.015514 \\
    Interface Engine/Protocol Translator &    146 &  0.012112 \\
    Electronic Design Automation (EDA)   &     92 &  0.007632 \\
    Artificial Life                      &     41 &  0.003401 \\
    \hline
    \end{tabular}
    \caption{Number of assigned classifications and proportion of total by Scientific/Engineering subtopic}
    \label{tab:stats_by_topic_scientific}
\end{table}

Finally, we examine the 156.8 million source code imports identified in the procedure described in Section \ref{S:dependency_identification_procedure} and Figure \ref{fig:dependency_identification_flowchart} above.  While the PyPI ecosystem may grow in terms of authors, packages, and releases, another important aspect is its degree of connectivity, both to standard library packages and to other PyPI packages.  A complete network- or graph-based treatment and analysis of PyPI is outside of the scope of this paper, but simple statistics on the number of import statements over time, i.e., edges, and number of imports per package/module, i.e., vertex indegree, is provided for summary here.

Firstly, we examine the number of imports over time as measured by year of release.  This measure is analogous to the number of new edges added to the network of software dependencies each year, or the rate of growth of the network.  Table \ref{tab:stats_import_by_year} documents this time series, demonstrating a 61.7\% compound annual growth rate in the number of imports, exceeding the growth rates of the number of packages or releases alone by 10\%.  This observation suggests that PyPI is growing not just in volume (number of vertices/packages), but also potentially in density (number of imports/edges); that is, the number of connections or imports \textit{per package} is increasing.  We examine actual dependency graph  statistics in future research, including the degree to which these imports are cross- or intra-package and cross- or intra-license.

\begin{table}[ht!]
    \centering
    \begin{tabular}{r|r}
    \hline
    \textbf{Year} & \textbf{Count} \\
    \hline
    2005 &     26,756 \\
    2006 &     91,896 \\
    2007 &    188,050 \\
    2008 &    459,461 \\
    2009 &   1,015,681 \\
    2010 &   1,520,202 \\
    2011 &   2,335,051 \\
    2012 &   3,552,860 \\
    2013 &   6,837,532 \\
    2014 &  10,864,275 \\
    2015 &  15,716,157 \\
    2016 &  25,696,479 \\
    2017 &  36,932,209 \\
    2018 &  47,745,271 \\
    2019 &   3,834,870 \\
    \end{tabular}
    \caption{Number of imports by year of release}
    \label{tab:stats_import_by_year}
\end{table}

Secondly, we analyze how these imports are distributed across specific packages and import strings.  Tables \ref{tab:stats_import_by_package} and \ref{tab:stats_import_by_import} below summarize these distributions by displaying the top 20 packages and import strings, including both raw count and proportion of total imports.  Table \ref{tab:stats_import_by_package} first documents the relative package frequency; unexpectedly, the highest frequency package import in all of PyPI is not a standard library package, but instead django.  Of the top 20 packages, nine of the packages, or nearly half, are not from the standard library, including plenum, homeassistant, numpy, pyangbind, zope, ccxt, flexget, setuptools, and tendenci.

\begin{table}[ht!]
    \centering
    \begin{tabular}{r|r|r}
    \hline
    \textbf{Package} & \textbf{Count} & \textbf{Proportion} \\
    \hline
    django        &  8,873,327 &    0.056584 \\
    \_\_future\_\_    &  6,116,026 &    0.039001 \\
    os            &  5,606,409 &    0.035751 \\
    sys           &  3,369,534 &    0.021487 \\
    logging       &  3,136,004 &    0.019998 \\
    re            &  2,452,253 &    0.015638 \\
    plenum        &  2,336,165 &    0.014897 \\
    datetime      &  1,935,426 &    0.012342 \\
    numpy         &  1,895,266 &    0.012086 \\
    json          &  1,693,092 &    0.010797 \\
    unittest      &  1,624,459 &    0.010359 \\
    homeassistant &  1,595,625 &    0.010175 \\
    time          &  1,507,953 &    0.009616 \\
    pyangbind     &  1,491,086 &    0.009508 \\
    zope          &  1,404,640 &    0.008957 \\
    ccxt          &  1,373,149 &    0.008756 \\
    setuptools    &  1,296,754 &    0.008269 \\
    flexget       &  1,283,930 &    0.008187 \\
    collections   &  1,233,227 &    0.007864 \\
    tendenci      &  1,226,318 &    0.007820 \\
    \end{tabular}
    \caption{Number of imports and proportion of total by package for top 20 packages}
    \label{tab:stats_import_by_package}
\end{table}

Many of these imports are likely due to high-frequency release strategies, automated source code generation for frameworks, object-relational mappers (ORM), library bindings, or testing suites.  When viewed through the lens of unique packages importing per year, many of those packages have much more reasonable counts, and the standard library packages are shown to be more broadly used across packages.  Table \ref{tab:stats_by_unique_package_import} shows the time series of unique importing packages for a sample of common top-level package imports, including both standard library and non-standard library packages.  The full time series data is available for all packages in the data repository referenced above in Section \ref{S:data}.

\begin{table}[ht!]
    \centering
    \begin{tabular}{r|r|r|r|r|r|r}
    \hline
    \textbf{Year} & \textbf{os} &    \textbf{sys} &     \textbf{re} &  \textbf{django} &  \textbf{ccxt} &  \textbf{numpy} \\
    \hline
    2005 & 47 &     55 &     29 &       0 &     0 &      0 \\
   2006 & 176 &    194 &     96 &       0 &     0 &     15 \\
   2007 & 556 &    457 &    221 &       7 &     0 &     20 \\
  2008 & 1339 &    892 &    503 &      38 &     0 &     55 \\
  2009 & 2439 &   1552 &    872 &     163 &     0 &    115 \\
  2010 & 3523 &   2482 &   1423 &     404 &     0 &    176 \\
  2011 & 4660 &   3341 &   2017 &     655 &     0 &    265 \\
  2012 & 6935 &   4896 &   3187 &    1032 &     0 &    474 \\
 2013 & 10111 &   7725 &   5053 &    1464 &     0 &    856 \\
 2014 & 13803 &  11159 &   7386 &    1763 &     0 &   1504 \\
 2015 & 18687 &  15131 &  10234 &    2068 &     0 &   2435 \\
 2016 & 23436 &  18944 &  13170 &    2290 &     0 &   3680 \\
 2017 & 29060 &  23108 &  16153 &    2247 &     1 &   5876 \\
 2018 & 37225 &  29185 &  20456 &    2330 &    15 &   9238 \\
  2019 & 9983 &   7209 &   5069 &     324 &     5 &   3013 \\
    \end{tabular}
    \caption{Number of unique importing packages by top-level package import}
    \label{tab:stats_by_unique_package_import}
\end{table}

We also examine imports below the top-level package, summarizing in Table \ref{tab:stats_import_by_import} the frequency of the top 20 import strings, including the proportion of total imports that they account for.  From this perspective, the standard libraries account for the top seven imports and 12 of the top 20 imports.  The specific import paths used for many of these packages also make clear their source as boilerplate code generation, such as the django.db or pyangbind.lib.yangtypes records.  

\begin{table}[ht!]
    \centering
    \begin{tabular}{r|r|r}
    \hline
    \textbf{Import String} & \textbf{Count} & \textbf{Proportion} \\
    \hline
    \_\_future\_\_              &  6,116,026 &    0.039001 \\
    os                      &  4,925,772 &    0.031411 \\
    sys                     &  3,369,515 &    0.021487 \\
    logging                 &  3,068,348 &    0.019566 \\
    re                      &  2,452,115 &    0.015637 \\
    datetime                &  1,935,350 &    0.012341 \\
    json                    &  1,688,541 &    0.010768 \\
    numpy                   &  1,658,696 &    0.010577 \\
    unittest                &  1,534,341 &    0.009784 \\
    time                    &  1,507,805 &    0.009615 \\
    collections             &  1,215,826 &    0.007753 \\
    django.db               &  1,152,455 &    0.007349 \\
    setuptools              &  1,080,083 &    0.006888 \\
    six                     &   923,327 &    0.005888 \\
    pytest                  &   873,933 &    0.005573 \\
    ccxt.base.errors        &   805,028 &    0.005134 \\
    utils                   &   780,507 &    0.004977 \\
    django.conf             &   754,178 &    0.004809 \\
    pyangbind.lib.yangtypes &   752,558 &    0.004799 \\
    os.path                 &   680,360 &    0.004339 \\
    \end{tabular}
    \caption{Number of imports and proportion of total by import string for top 20 imports}
    \label{tab:stats_import_by_import}
\end{table}

From a licensing and information security perspective, each import statement on PyPI is a path that requires traversing to ensure compliance and safety.  Luckily, many of the millions of import statements on PyPI are simple references to standard library modules or repetitive imports to the same packages or modules.

\section{Conclusion and Future Work}
\label{S:conclusion}

Motivated by gaps in the focus and data quality of extant research, we examine the package metadata and source code for the Python Package Repository (PyPI).  Our data expands on and improves that available through PyPI and other sources, covering over 175,000 packages, 1.7 million releases, 75,000 authors, and 150 million import statements as of May 2019.  While our analysis is generally of a summary nature, we cover a broad range of topics, such as the number of active and new packages and authors, the types and activity levels of authors contributing, the inter-release timing distribution, the size of packages and releases, the relative frequency of common open source licenses, and the assignment of common organizational classifications. Furthermore, we take a novel approach to the high-level analysis of the dependency network within PyPI by analyzing actual import statements at the file level instead of relying solely on often incorrect or incomplete package metadata

In future work, the authors intend to address a number of topics, including license and code re-use identification methodologies that handle the complexities of real software, problematic chains of license dependency, network analysis of the complete dependency graph, and an analysis of author and package demographics and dynamics, including not just package metadata and point-in-time source, but also VCS history where available.

\section{Acknowledgements}
\label{S:acknowledgements}

The authors would like to acknowledge the late Professor Rick Riolo, whose spirit of inquiry into complex systems everywhere lives on in his many students around the world.

%% References


\bibliographystyle{model1-num-names}
\bibliography{empirical_python.bib}






% \end{thebibliography}

\end{document}